\documentclass[%
 reprint,
superscriptaddress,
 amsmath,amssymb,
 aps,
 pra,
]{revtex4-1}

\usepackage{xcolor} 
\usepackage[version=3]{mhchem} 
\usepackage{graphicx}
\usepackage{dcolumn}
\usepackage{bm}
\usepackage{tabularx}
\usepackage{rotating}
\usepackage{makecell}

\begin{document}

\title{Structure and magnetism of the skyrmion hosting family \ce{GaV4S_{8-y}Se_y} with low levels of substitutions between $0 \leq y \leq 0.5$ and  $7.5 \leq  y\leq 8$}

\author{S. J. R. Holt}
\email{S.J.R.Holt@warwick.ac.uk}
\affiliation{University of Warwick, Department of Physics, Coventry, CV4 7AL, United Kingdom}
\author{A. \v{S}tefan\v{c}i\v{c}}
\altaffiliation{Present address: Electrochemistry Laboratory, Paul Scherrer Institut, CH-5232 Villigen PSI, Switzerland}
\affiliation{University of Warwick, Department of Physics, Coventry, CV4 7AL, United Kingdom}
\author{C. Ritter}
\affiliation{Institut Laue-Langevin, 71 Avenue des Martyrs, CS20156, 38042 Grenoble C\'edex 9, France}
\author{A. E. Hall}
\affiliation{University of Warwick, Department of Physics, Coventry, CV4 7AL, United Kingdom}
\author{M. R. Lees}
\affiliation{University of Warwick, Department of Physics, Coventry, CV4 7AL, United Kingdom}
\author{G. Balakrishnan}
\email{G.Balakrishnan@warwick.ac.uk}
\affiliation{University of Warwick, Department of Physics, Coventry, CV4 7AL, United Kingdom}

\date{\today}

\begin{abstract}
Polycrystalline members of the \ce{GaV4S_{8-y}Se_y} family of materials with small levels of substitution between $0 \leq y \leq 0.5$ and  $7.5 \leq  y\leq 8$  have been synthesized in order to investigate their magnetic and structural properties.
Substitutions to the skyrmion hosting parent compounds \ce{GaV4S8} and \ce{GaV4Se8}, are found to suppress the temperature of the cubic to rhombohedral structural phase transition that occurs in both end compounds and to create a temperature region around the transition where there is a coexistence of these two phases.
Similarly, the magnitude of the magnetization and temperature of the magnetic transition are both suppressed in all substituted compounds until a glassy-like magnetic state is realized.
There is evidence from the \textit{ac} susceptibility data that skyrmion lattices with similar dynamics to those in \ce{GaV4S8} and \ce{GaV4Se8} are present in compounds with very low levels of substitution, $0 < y< 0.2$ and  $7.8 < y < 8$, however, these states vanish at higher levels of substitution.
The magnetic properties of these substituted materials are affected by the substitution altering exchange pathways and resulting in the creation of increasingly disordered magnetic states.
\end{abstract}

\maketitle

\section{\label{sec:Intro} Introduction}
In recent years there has been significant interest in the field of magnetic skyrmions and in the search for new materials which exhibit skyrmions \cite{lancaster2019skyrmions}.  
While an increasing number of materials that host bulk Bloch skyrmions have been discovered, to date there are only a limited number of known bulk systems in which N\'eel skyrmions are stabilized, hence further investigation into the discovery of these systems is of great interest \cite{kezsmarki2015neel, fujima2017thermodynamically, kurumaji2017neel, schueller2020structural, bordacs2017equilibrium}.
One way to increase the number of skyrmion hosting materials is to extend the families of known skyrmion materials by chemical manipulation.
For example, in materials such as \ce{Cu2OSeO3} the substitution at the Cu site has enabled the tuning of physical parameters in the system, and hence their magnetic properties \cite{stefancic2018origin, wilson2019measuring, birch2019increased}.
In the search for new skyrmion hosts it is also necessary to gain a detailed understanding of the fundamental links between the crystal structures adopted by skyrmionic materials and the magnetic states that they exhibit, to understand what makes these materials suitable hosts. 

One family of materials in which a few members have been shown to stabilize N\'eel type skyrmions (for example \ce{GaV4S8} and \ce{GaV4Se8}) are the lacunar spinels $AB_4X_8$ ($A$~=~Ga, Al, Ge, Ti, Fe, Co, Ni, Zn; $B$~=~V, Cr, Mo, Re, Nb, Ta; $X$~=~S, Se, Te), which have been studied due to their intriguing electronic and magnetic properties \cite{barz1973new, perrin1975new, brasen1975magnetic, perrin1976nouveaux, yaich1984nouveaux, rastogi1984electron, johrendt1998crystal, pocha2000electronic, pocha2005crystal, bichler2007tuning, powell2007cation, vaju2008metal, szkoda2009compositional, bichler2011interplay}.
Lacunar spinels undergo structural phase transitions at low temperatures giving rise to a variety of magnetic phases exhibiting a plethora of interesting properties, such as skyrmions. 
Of these, \ce{GaV4S8} and \ce{GaV4Se8} have been extensively studied as skyrmion hosts \cite{kezsmarki2015neel, fujima2017thermodynamically, stefancic2020establishing, bordacs2017equilibrium}.
Both \ce{GaV4S8} and \ce{GaV4Se8} are isostructural with a structural phase transition from a high temperature cubic $F\Bar{4}3m$ phase to a low temperature rhombohedral $R3m$ structure at $42$ and $43$~K, respectively with a further magnetic transition at 13 and 18~K, respectively \cite{pocha2000electronic, bichler2011interplay}.
The high temperature cubic structure consists of V$_4X_4$ and Ga$X_4$ tetrahedra arranged in a \ce{NaCl} structure.
In this structure, there are two distinct sites on which the chalcogenides can sit: either the S1/Se1 site within the V$_4X_4$ cluster or the S2/Se2 site within the Ga$X_4$ tetrahedra.
It has been suggested that the cubic to rhombohedral phase transition is driven by such mechanisms as a cooperative Jahn-Teller distortion \cite{wang2015polar} and charge order of the \ce{V4} cluster formed of three \ce{V^{3+}} and one \ce{V^{4+}} ion \cite{stefancic2020establishing}.
In the low temperature rhombohedral phase a single V ion distorts along the one of the four possible cubic $\langle 1~1~1 \rangle$ directions in the the V$_4X_4$ cluster.
This distortion causes a reduction in symmetry and removes the triple degeneracy of the $t_2$ orbitals splitting them into two higher energy $e$ levels and one lower energy $a_1$ level. 
One other related member of the lacunar spinels, \ce{GaMo4Se8}, has recently been reported to exhibit \cite{schueller2020structural} large changes in the magnetic phase diagram brought about as a result of subtle structural distortions.
\ce{GaMo4Se8}, which adopts a cubic $F\Bar{4}3m$ structure at room temperature is reported to undergo a transition at low temperatures to a ground state where there is a coexistence of two distinct structural space groups, a $R3m$ phase and a metastable $Imm2$ phase, exhibiting distinctly different magnetic properties strongly related to their magneto-crystalline anisotropy.
In this material, the $R3m$ phase is known to exhibit N\'eel type skyrmions \cite{schueller2020structural} and is a clear demonstration of the tunability of these structures to effect dramatic changes in the magnetic properties.

\ce{GaV4S8} hosts a variety of magnetic phases.
In zero-field there is a transition from a paramagnetic to cycloidal magnetic state \cite{kezsmarki2015neel}.
Further reduction in temperature results in a commensurate magnetic state with the moments on the distorted V ion aligning ferromagnetically and the moments on the other three V ions canted in a $R3m'$ Shubnikov group pointing out of the cluster \cite{stefancic2020establishing}.
Application of a small magnetic field to the cycloidal phase stabilizes magnetic skyrmions, and further application of a magnetic field results in the formation a field polarized state \cite{kezsmarki2015neel}.
In comparison, \ce{GaV4Se8} shows a transition from a paramagnetic to a cycloidal state with decreasing temperature and remains cycloidal down to 1.5~K.
Application of a small magnetic field to the cycloidal phase is found to stabilize magnetic skyrmions followed by a field polarized state at higher fields~\cite{bordacs2017equilibrium}.
In both \ce{GaV4S8} and \ce{GaV4Se8} the skyrmions form along the [111] pseudo-cubic crystallographic direction while having strong easy axis and easy plane anisotropy respectively.
The extent of the skyrmion phase strongly depends on the orientation of applied magnetic field with respect to the crystallographic direction \cite{bordacs2017equilibrium}.
Muon-spin rotation ($\mu^+$SR) measurements have shown that there are large differences in the magnetic ground states of \ce{GaV4S8} and \ce{GaV4Se8}, with \ce{GaV4Se8} showing a peculiar increase in local magnetic fields with an increase of temperature \cite{hicken2020magnetism,franke2018magnetic}.

Given that both \ce{GaV4S8} and \ce{GaV4Se8} host skyrmions, we have investigated in a previous study the entire solid solution formed between these two parent materials, by substituting S with Se, in an attempt to see if any of the intermediate phases exhibit the pre-requisite magnetic and structural properties to host skyrmions \cite{stefancic2020establishing}.
This study revealed that the solid solutions formed for $1\leq y\leq 7$  (in integer intervals) remain cubic down to 1.5~K, thus preventing the formation of N\'eel type skyrmions.
In addition, it was found that these intermediate compositions do not exhibit magnetic ordering, and instead have glassy-like magnetic ground states indicative of magnetic disorder within the system.
This study highlights the important structure-magnetism correlations in these materials.
The substitution of S/Se, for example, in the two parent materials was not expected to produce a large disruption to the local structural units but this was not the case, as was evidenced.

In order to further examine how the evolution of the magnetic states is intricately linked to structural transformations and instabilities, we have now synthesized and investigated selected compositions very close to the two end compounds in the \ce{GaV4S_{8-y}Se_y} family of materials ($y=$ 0 to 0.5 and $y=$ 7.5 to 8).
By effecting extremely subtle distortions by these low level substitutions, we have been able to examine the stability of these local structural units, which are crucial to these materials acting as skyrmion hosts.
We have also been able to evaluate the critical S/Se substituted compositions at which the transition from a skyrmion hosting to a non-skyrmion hosting material occurs and determine the mechanisms leading to this. 
Detailed structural analysis has been carried out using both X-ray and neutron diffraction.
The magnetic properties have been investigated using \textit{ac}/\textit{dc} magnetometry to understand the strong correlations between the structure and magnetic behaviors existing in these systems and to probe the critical phenomena in this narrow composition range.
Our results indicate that small levels of substitution create a coexistence of both cubic $F\Bar{4}3m$ and rhombohedral $R3m$ structural phases in the region surrounding the temperature of the structural phase transition.
This phenomena is coupled with a reduction of the structural transition temperature with increasing substitution levels, until the structural transition is no longer present for $y=0.5$ and nearly eradicated for $y=7.5$.
The magnetic susceptibility is dramatically suppressed and the nature of the magnetic phase diagram is significantly altered with increasing levels of substituents.
We show that both the structural and magnetic properties of the substituted materials exhibit marked changes for substitution levels as low as $2.5\%$ (i.e. $0 \leq y<0.2$ and  $7.8 < y\leq 8$) in both parent compounds, thus demonstrating the fragility of the systems to substitutional or similar disorder and its effect on the ability of these systems to host skyrmions. 

\section{\label{sec:ExpDets} Experimental Details}
Polycrystalline \ce{GaV4S_{8-y}Se_y} samples with $y=$~0, 0.1, 0.2, 0.3, 0.4, 0.5, 7.5, 7.6, 7.7, 7.8, 7.9, and 8 were synthesized by the standard solid-state reaction method similar to that used for the synthesis of polycrystalline \ce{GaV4S8} and \ce{GaV4Se8}~\cite{stefancic2020establishing}.
Powder X-ray diffraction was performed on a Panalytical X-Pert Pro diffractometer operating in Bragg-Brentano geometry equipped with a monochromatic Cu K$_{\alpha 1}$ source and a solid-state PIXcel one-dimensional detector, to determine the phase purity and crystallographic structure of the polycrystaline materials at room temperature.
Rietveld refinements were carried out using the TOPAS academic v6.0 software refinement based on the structural model obtained by single crystal X-ray diffraction in our previous study \cite{stefancic2020establishing, Coelho2018TOPAS}.

Powder neutron diffraction was performed on the D2B diffractometer at Institut Laue Langevin (ILL) to establish the nuclear structures between $1.5$ and $50$~K.
Powder samples were sealed in thin-walled cylindrical vanadium containers of 8 to 12~mm diameter and placed in a standard orange cryostat.
A neutron wavelength of $1.594$~\AA~was used to obtain high-resolution diffractograms for nuclear structure refinements.
Structural Rietveld refinements have been used to simultaneously fit the diffractograms of a single composition whilst imposing physical behavior using TOPAS academic v6.0 \cite{Coelho2018TOPAS}.

The magnetic properties of the polycrystaline materials were measured using a Quantum Design Magnetic Property Measurement System, superconducting quantum interference device (SQuID) magnetometer.
\textit{dc} temperature-dependent magnetic susceptibility $(\chi)$ measurements were carried out at temperatures between 1.8 and 300~K in an applied field of 10~mT using a zero-field-cooled-warming (ZFCW) protocol.
\textit{ac} susceptibility measurements were performed with a 113~Hz \textit{ac} field of 0.3~mT after zero-field-cooling to each measuring temperature and then applying a \textit{dc} magnetic field.

\section{\label{sec:Results} Results and Discussion}
\subsection{\label{sec:PXRD} Powder X-ray diffraction}
\begin{figure}
\centering
\includegraphics[width=0.9\linewidth]{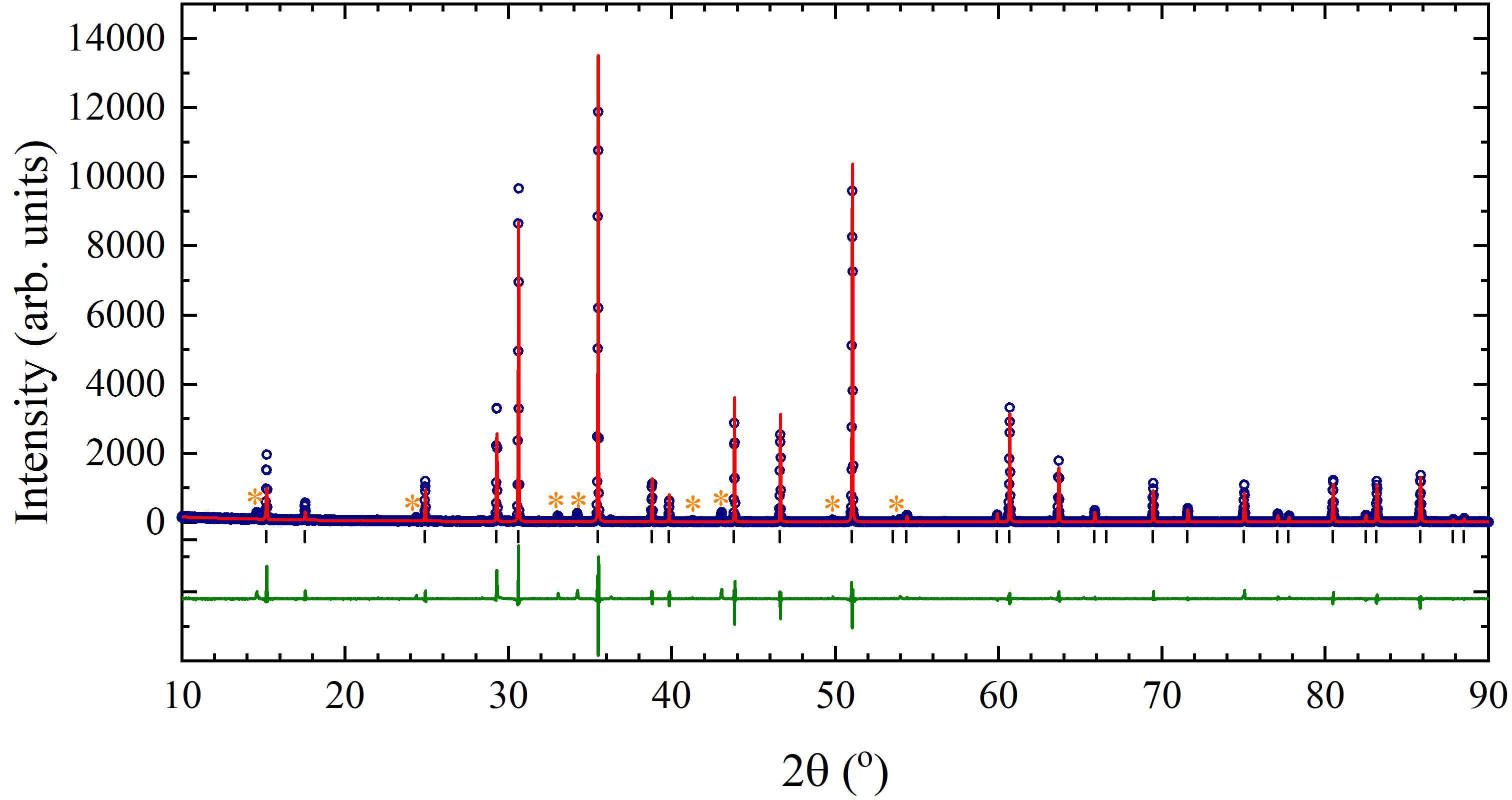}
\caption{Powder X-ray diffraction pattern of \ce{GaV4S_{0.4}Se_{7.6}}. Experimentally-obtained diffraction profile at ambient temperature (blue open circles), predicted peak positions (black tick marks), Rietveld refinement (red solid line), difference (olive green solid line). 
$R_\text{p} =$~15.4, $R_\text{wp}=$~21.3, and $R_\text{exp}=$~11.1.
* indicate small impurity peaks belonging to \ce{VSe2}.}
\label{Fig:PXRD}
\end{figure}

\begin{figure}
\centering
\includegraphics[width=0.9\linewidth]{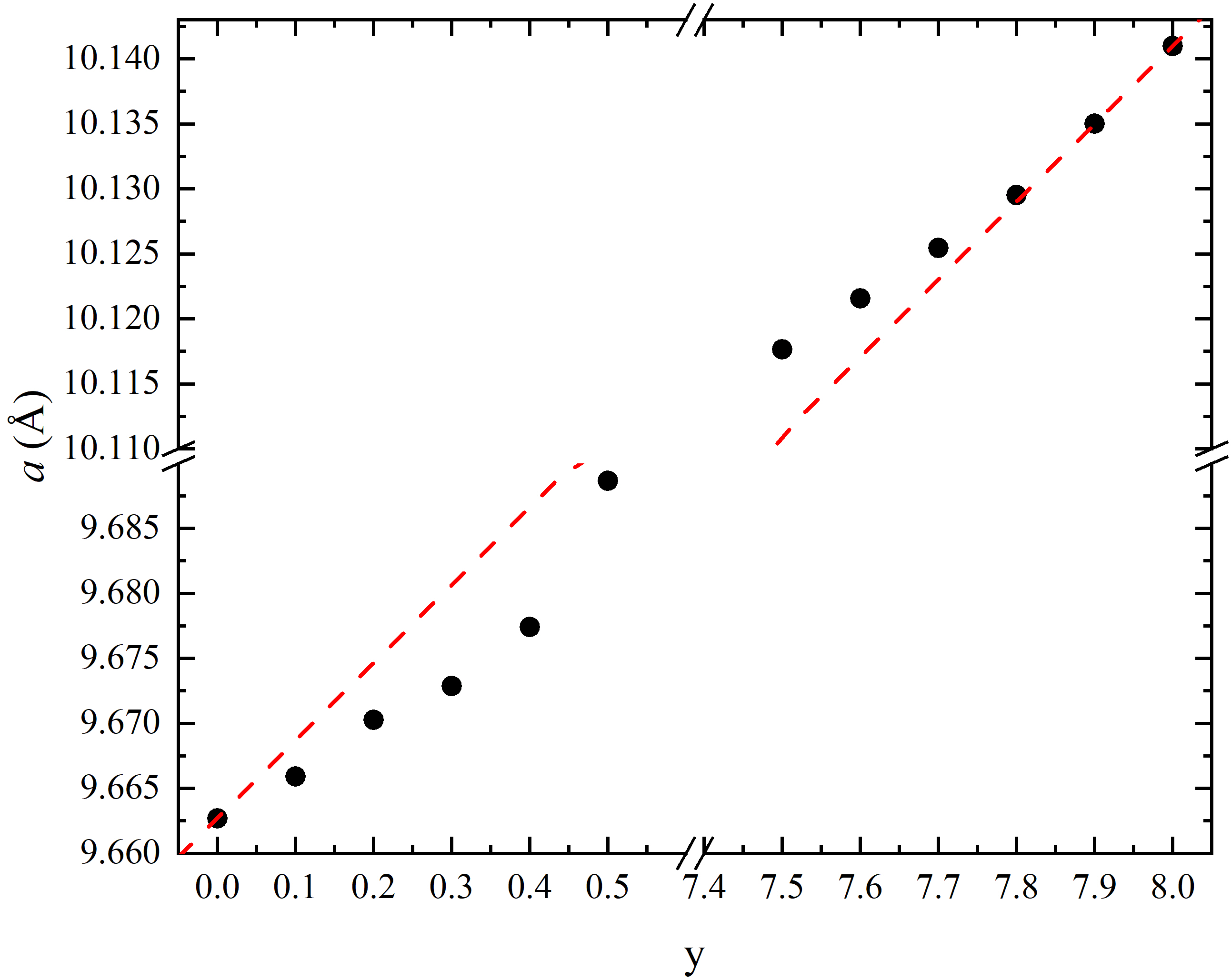}
\caption{Lattice parameters obtained from Rietveld refinements of powder X-ray diffraction patterns of GaV$_4$S$_{8-y}$Se$_{y}$ $0\leq y\leq 0.5$ and $7.5\leq y\leq 8$  at ambient temperature in the $F\bar{4}3m$ structural phase. The dashed red line represent Vegard's law and the error bars are too small to be seen.}
\label{Fig:Lat}
\end{figure}

Room temperature powder X-ray diffraction data of the polycrystaline \ce{GaV4S_{8-y}Se_y} with $0\leq y\leq 0.5$ and $7.5\leq y\leq 8$ could all be indexed in $F\bar{4}3m$ cubic symmetry.
A representative pattern is shown in Fig.~\ref{Fig:PXRD} for one of the materials investigated,  \ce{GaV4S_{0.4}Se_{7.6}}.
The patterns show that the materials form predominantly as single phase with low levels of impurities such as \ce{VSe2} for $7.5\leq y\leq 8$ and \ce{V5S8} for $0\leq y\leq 0.5$.
Figure~\ref{Fig:Lat} indicates a monotonically increasing lattice parameter with increasing Se content which was also observed in our earlier study covering a much wider substitution range~\cite{stefancic2020establishing}.

\subsection{\label{sec:PND} Powder Neutron Diffraction}
The structural properties of the family have been examined using high resolution powder neutron diffraction on the D2B diffractometer at the ILL.

\subsubsection{\label{sec:SPT} Structural phase transition}

\begin{figure}[t]
\centering
\includegraphics[width=0.9\linewidth]{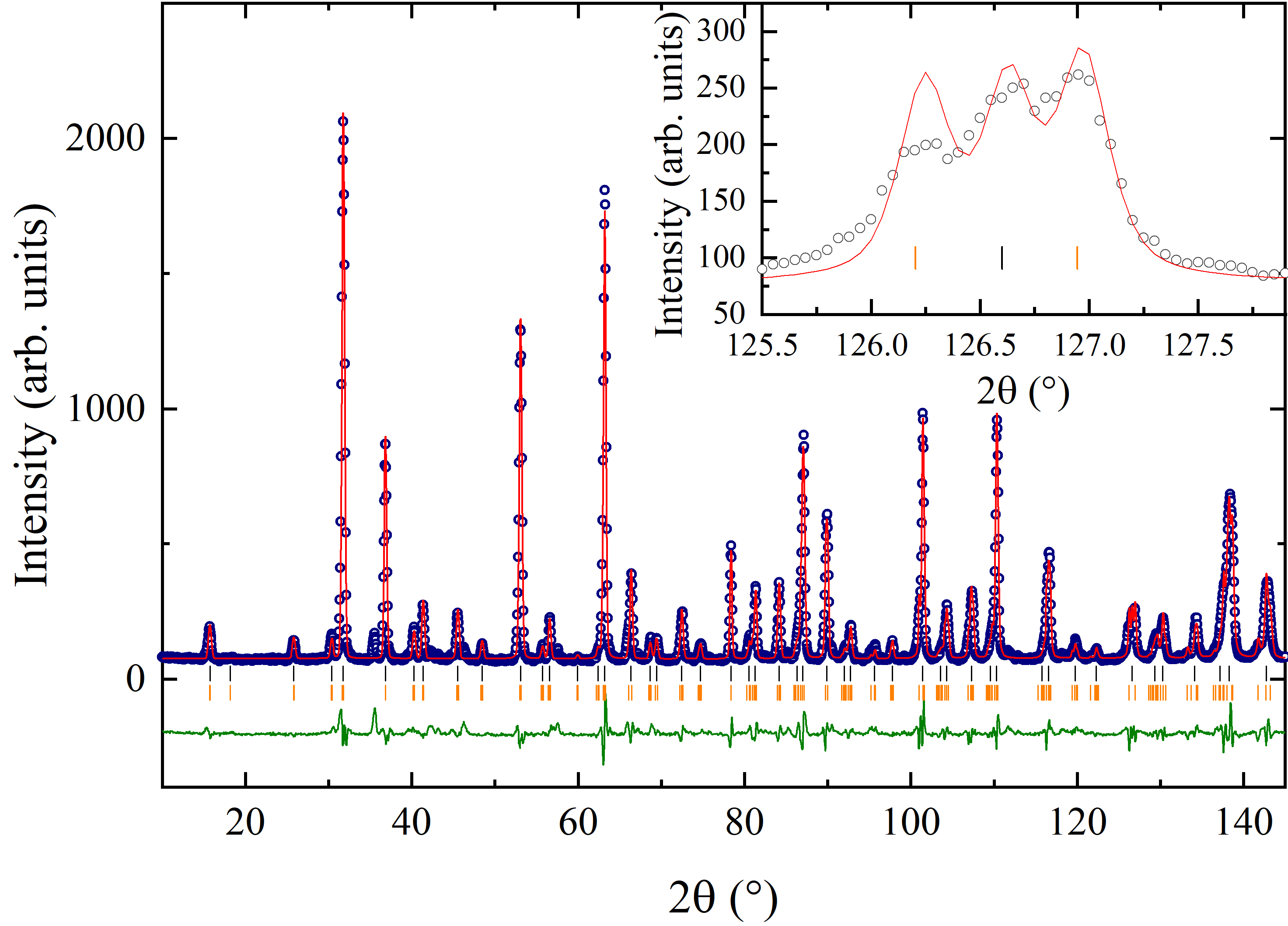}
\caption{Experimentally obtained neutron diffraction profile at 1.5~K (blue open circles) for \ce{GaV4S_{0.5}Se_{7.5}}, Rietveld refinements (red solid line), difference (olive green solid line), predicted peak positions of the $F\bar{4}3m$ (black tick marks), and predicted peak positions of the $R3m$ (orange tick marks).
$R_\text{p} =$~6.5, $R_\text{wp}=$~8.5, and $R_\text{exp}=$~4.8.
The inset shows the coexistence of the $(8~8~0)$ $F\bar{4}3m$ diffraction peak with the hexagonal $R3m$ $(4~0~16)$ and $(8~\bar{4}~0)$ peaks at 1.5~K.}
\label{Fig:D2B_Fit}
\end{figure}

\begin{figure}[t]
\centering
\includegraphics[width=0.9\linewidth]{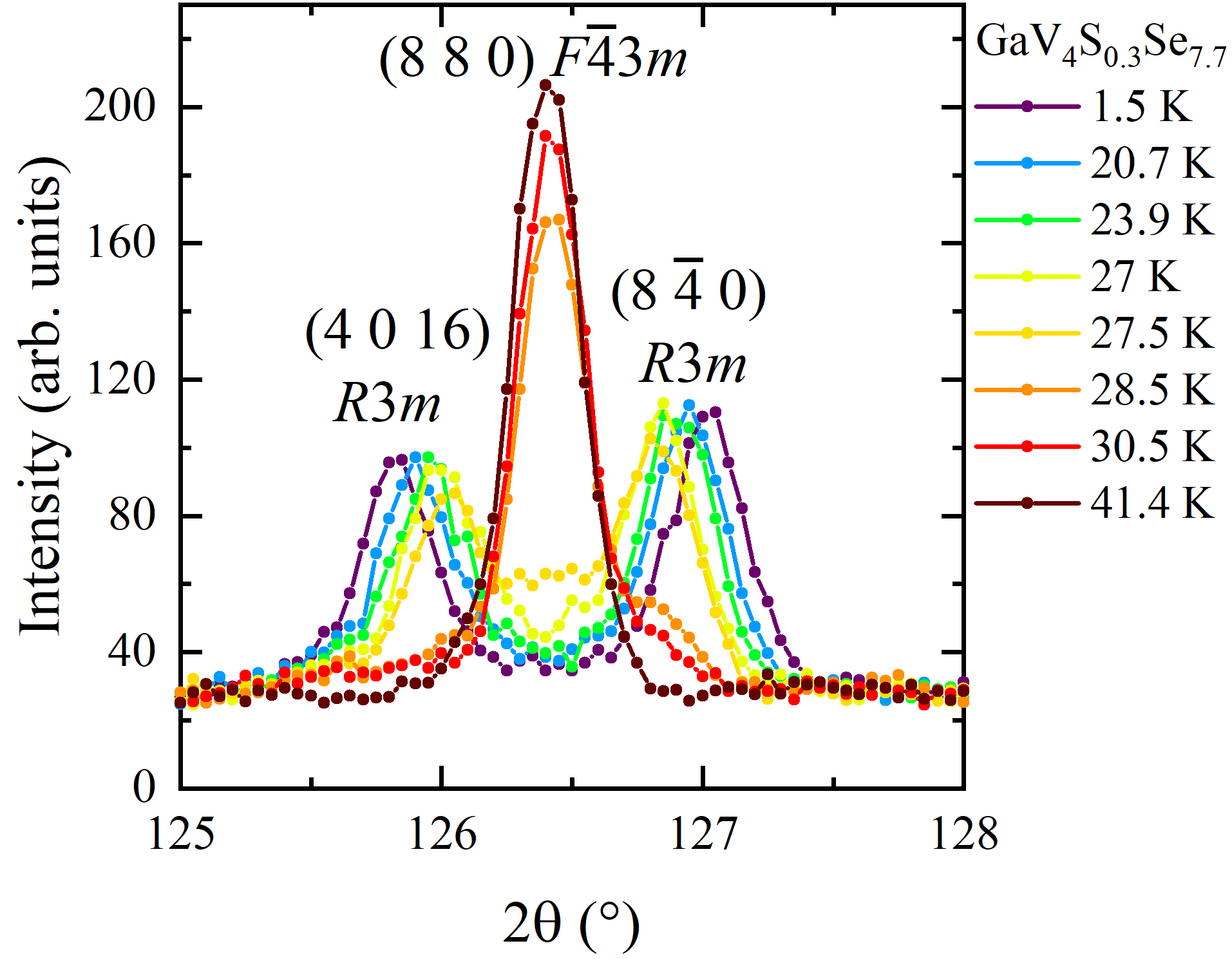}
\caption{High-resolution neutron diffractograms taken using the D2B diffractometer at the ILL show splitting of the cubic $(8~8~0)$ $F\bar{4}3m$ Bragg peak for \ce{GaV4S_{0.3}Se_{7.7}} into the hexagonal $R3m$ $(4~0~16)$ and $(8~\bar{4}~0)$ Bragg peaks in the temperature range 1.5 to 41.4~K.}
\label{Fig:D2B_Split}
\end{figure}

Both \ce{GaV4S8} and \ce{GaV4Se8} were found to undergo a structural phase transition from a high temperature cubic $F\Bar{4}3m$ phase to a lower temperature rhombohedral $R3m$ structure, as previously reported~\cite{pocha2000electronic, bichler2007tuning}.
The Rietveld refinements performed on the parent and substituted GaV$_4$S$_{8-y}$Se$_{y}$ compounds with $0\leq y\leq 0.1$, $0.4\leq y\leq 0.5$, and $7.5\leq y\leq 8$ reveal some interesting structural properties of the family including a region of structural phase coexistence.
As an example, a typical Rietveld refinement of the high resolution neutron diffraction data collected on the $y=7.5$ sample is shown in Fig.~\ref{Fig:D2B_Fit}.
Here, the results of the refinements reveal that there is a coexistence of two phases having $F\bar{4}3m$ and $R3m$ symmetry.
This is seen as a splitting of the cubic $(8~8~0)$ peak depicted in the inset of Fig.~\ref{Fig:D2B_Fit} for the refinement at 1.5~K where the coexistence of both cubic and rhombohedral peaks is evident.
The lower temperature rhombohedral $R3m$ structure is pseudo-cubic, which results in the nuclear diffraction peaks being close to those of the high temperature cubic $F\Bar{4}3m$ phase.
Hence, the structural phase transition is best revealed as a splitting of high order cubic Bragg peaks.
This is illustrated in Fig.~\ref{Fig:D2B_Split} for \ce{GaV4S_{0.3}Se_{7.7}} and is similar to what is observed in all of the compounds that undergo a structural phase transition ($0\leq y\leq 0.4$ and $7.5\leq y\leq 8$).  
Figure~\ref{Fig:D2B_Split} also shows the shift of the angle at which the $R3m$ Bragg peaks occur, illustrating the thermal expansion of the $c$ axis and contraction of the $a$ axis, reflecting the increasing rhombohedral distortion.

\begin{figure}[t]
\centering
\includegraphics[width=0.9\linewidth]{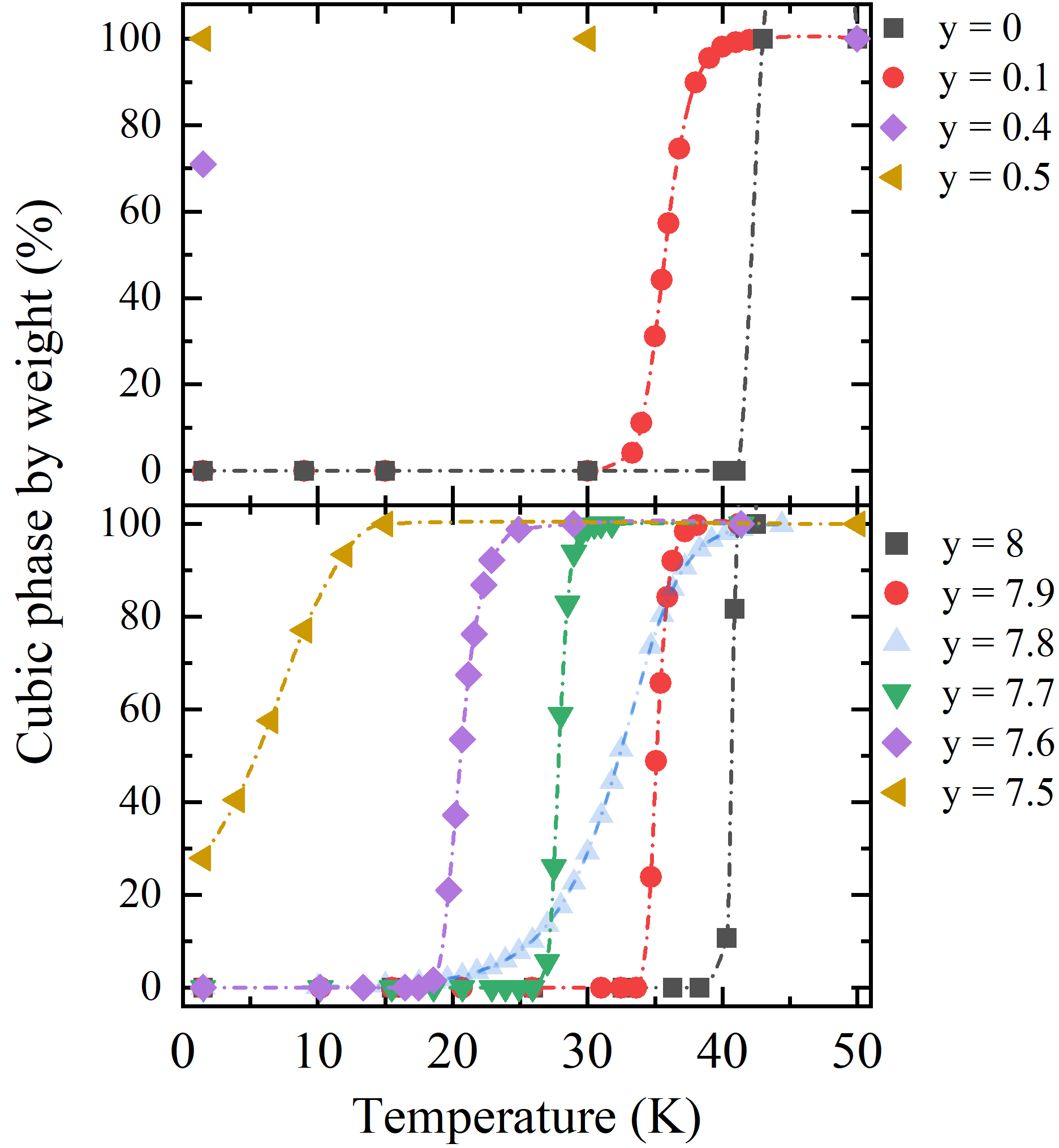}
\caption{Structural phase transition from cubic $F\Bar{4}3m$ phase to a rhombohedral $R3m$ by quantitative phase analysis of high-resolution neutron diffractograms collected on the D2B diffractometer at the ILL using Rietveld refinements of GaV$_4$S$_{8-y}$Se$_{y}$ powders. Top: $y= 0$, 0.1, 0.4, and 0.5. Bottom: $y= 8$, 7.9, 7.8, 7.7, 7.6, and 7.5.}
\label{Fig:Phase_weight}
\end{figure}

Figure~\ref{Fig:Phase_weight} depicts the observed structural phase transitions quantified as the percentage remainder of the cubic phase as a function of temperature and composition.
It can be seen that the two end phases, \ce{GaV4S8} and \ce{GaV4Se8}, undergo, as expected, a sharp structural phase transition, from $100\%$ $F\Bar{4}3m$ phase to $100\%$ $R3m$ phase at $42$ and $43$~K, respectively.
The sharp transitions for these two parent compounds indicate the first-order nature of their transitions, in agreement with literature \cite{pocha2000electronic, fujima2017thermodynamically}.
Figure~\ref{Fig:Phase_weight} shows that at both the S and Se ends of the series, the addition of substituents suppresses the onset temperature of the structural phase transition whilst simultaneously broadening it.
This broadening of the transition creates a wide temperature region in which refinements show the presence of both the $F\Bar{4}3m$ and $R3m$ structural phases.
Indication of the phase coexistence comes from features such as the appearance of a shoulder on the right side of the cubic $(8~8~0)$ peak in Fig.~\ref{Fig:D2B_Split} at 30.5 and 28.5~K.
Additionally, at lower temperatures there is still some intensity from the cubic $(8~8~0)$ peak located between the $(4~0~16)$ and $(8~\bar{4}~0)$ rhombohedral peaks.
Broadening of the structural phase transition is indicative of a disorder-broadened first-order phase transition \cite{roy2013first, uehara1999percolative}.
A certain amount of broadening is likely to be inherent to a particular substitution level as small levels of substitution preclude the possibility of homogeneity across all unit cells in the material. 
The extra broadening of the phase coexistence region observed for $y=7.8$ in Fig.~\ref{Fig:Phase_weight} compared to the other compositions investigated is most likely related to an increase of the variance of the local substitution level within the sample of this composition.


For \ce{GaV4S_{8-y}Se_y} with $7.5\leq y\leq 8$, Fig.~\ref{Fig:Phase_weight} shows a clear decrease of the transition temperature with the addition of substituents.
The region of cubic and rhombohedral coexistence for \ce{GaV4S_{0.5}Se_{7.5}} extends to low temperatures and the structural phase transition is incomplete with a coexistence of the two structural phases observed down to 1.5~K.
For the S rich materials, $y=0.5$ remains cubic down to $1.5$~K and $y=0.4$ exhibits a coexistence of the $F\bar{4}3m$ and $R3m$ phase at 1.5~K.
A similar trend of the decrease of the transition temperature with the addition of substituents is seen for compositions in the range $0\leq y\leq 0.5$, conforming to what is expected for higher substitution levels ($1\leq y\leq 7$), all of which remain cubic~\cite{stefancic2020establishing}.

In our previous study \cite{stefancic2020establishing} we suggested a couple of possible mechanisms that could be responsible for the suppression of the structural phase transition in this family of materials. These are the uneven local distribution of S/Se compared to neighboring unit cells creating a large elastic energy cost for the structural distortion, and the S/Se disorder within the sample changing electron diffusion pathways and preventing the structural phase transition. 
 
A similar observation has been made by Powell \textit{et al}. \cite{powell2007cation} in their investigations of the related GaV$_{4-x}$Mo$_x$S$_{8}$ family of materials, where  there is a  suppression of the structural phase transition in the intermediate materials, with these samples remaining in the cubic phase down to low temperatures. 
They suggest that this is due to the lifting of the electron degeneracy on a local level within the substituted \ce{V4S4} clusters, a mechanism which is equally applicable to the GaV$_4$S$_{8-y}$Se$_{y}$ series. 
The structural phase transition usually lifts the electron degeneracy upon cooling in the end compositions, but if the addition of substituents removes the electron degeneracy instead, then it could suppress the phase transition.
The consequences of the substitution related to the magnetic behavior are addressed in Sec.~\ref{sec:Mag}.
Our observations of the coexistence of two structural phases in GaV$_4$S$_{8-y}$Se$_{y}$ are very similar to the observations of  Powell \textit{et al.} in some of the intermediate GaV$_{4-x}$Mo$_x$S$_{8}$ materials \cite{powell2007cation}, where they also report the coexistence of $F\bar{4}3m$ and $R3m$ structural phases in some of the intermediate compositions investigated.  


\subsubsection{\label{sec:dis} Structural Distortion}
\begin{figure}[t]
\centering
\includegraphics[width=0.9\linewidth]{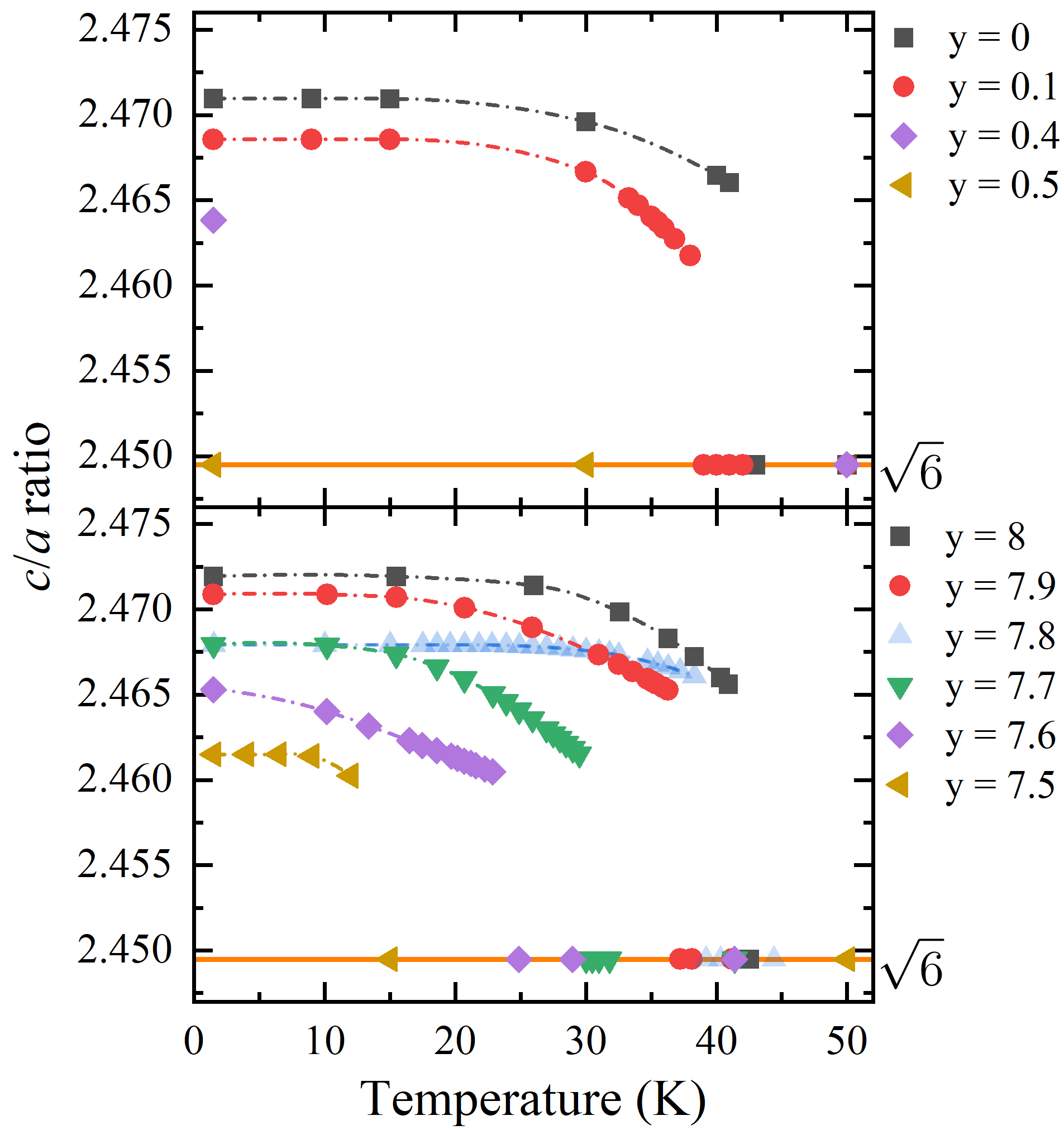}
\caption{Structural distortion of the rhombohedral $R3m$ crystal structure with respect to the cubic $F\Bar{4}3m$ structure in GaV$_4$S$_{8-y}$Se$_{y}$ powders as determined by Rietveld refinements of high-resolution neutron diffraction profiles taken using the D2B diffractometer at the ILL. The orange lines represent the constant value of the ratio for the cubic phase, which takes the value of $\sqrt{6}$. Top: $y= 0$, 0.1, 0.4, and 0.5. Bottom: $y= 8$, 7.9, 7.8, 7.7, 7.6, and 7.5.}
\label{Fig:Distortion}
\end{figure}

The distortion of the lattice when transforming from the cubic to the rhombohedral phase can be quantified by examining the $c/a$ lattice parameter ratio of the $R3m$ phase in the hexagonal setting.
This ratio takes a constant value of $\sqrt6$ as the rhombohedral distortion goes to zero.
The magnitude of the distortion as a function of temperature observed in the \ce{GaV4S_{8-y}Se_y} powders examined is plotted in Fig.~\ref{Fig:Distortion}.
The addition of substituents in both the S and Se ends of the family of materials is found to reduce the distortion of the $R3m$ phase.
As the temperature is increased, all the compositions examined show a similar gradual decrease in the amount of distortion followed by a sharp discontinuity at the rhombohedral to cubic phase transition temperature.
This discontinuity in the distortion is further evidence for the first-order nature of the structural transition.

Rietveld refinements of the neutron data confirm the preferential site occupation of Se in the Ga$X_4$ tetrahedra and S in the V$_4X_4$ clusters as expected for this system~\cite{stefancic2020establishing}. 
This occupational preference is attributed to the larger Se$^{2-}$ preferring the tetrahedral coordination of the Ga$X_4$ tetrahedra.
The improved statistics obtained with powder neutron diffraction over that obtained in the laboratory X-ray diffraction data, allows the more accurate refinement/determination of the preferential site occupancy, even for low levels of substitutions.
For GaV$_4$S$_{8-y}$Se$_{y}$ with $0< y\leq 0.5$, all the substituted Se was observed to occupy Ga$X_4$ tetrahedra, and all substituted S in $7.5\leq y<8$ was observed to occupy V$_4X_4$ clusters.
When undergoing the structural phase transition from the cubic to rhombohedral phase, the V$_4X_4$ clusters can distort along any of four possible cubic $\langle 1~1~1 \rangle$ directions.

\subsection{\label{sec:Mag} \textit{dc} and \textit{ac} magnetization}
\begin{figure}
\centering
\includegraphics[width=0.9\linewidth]{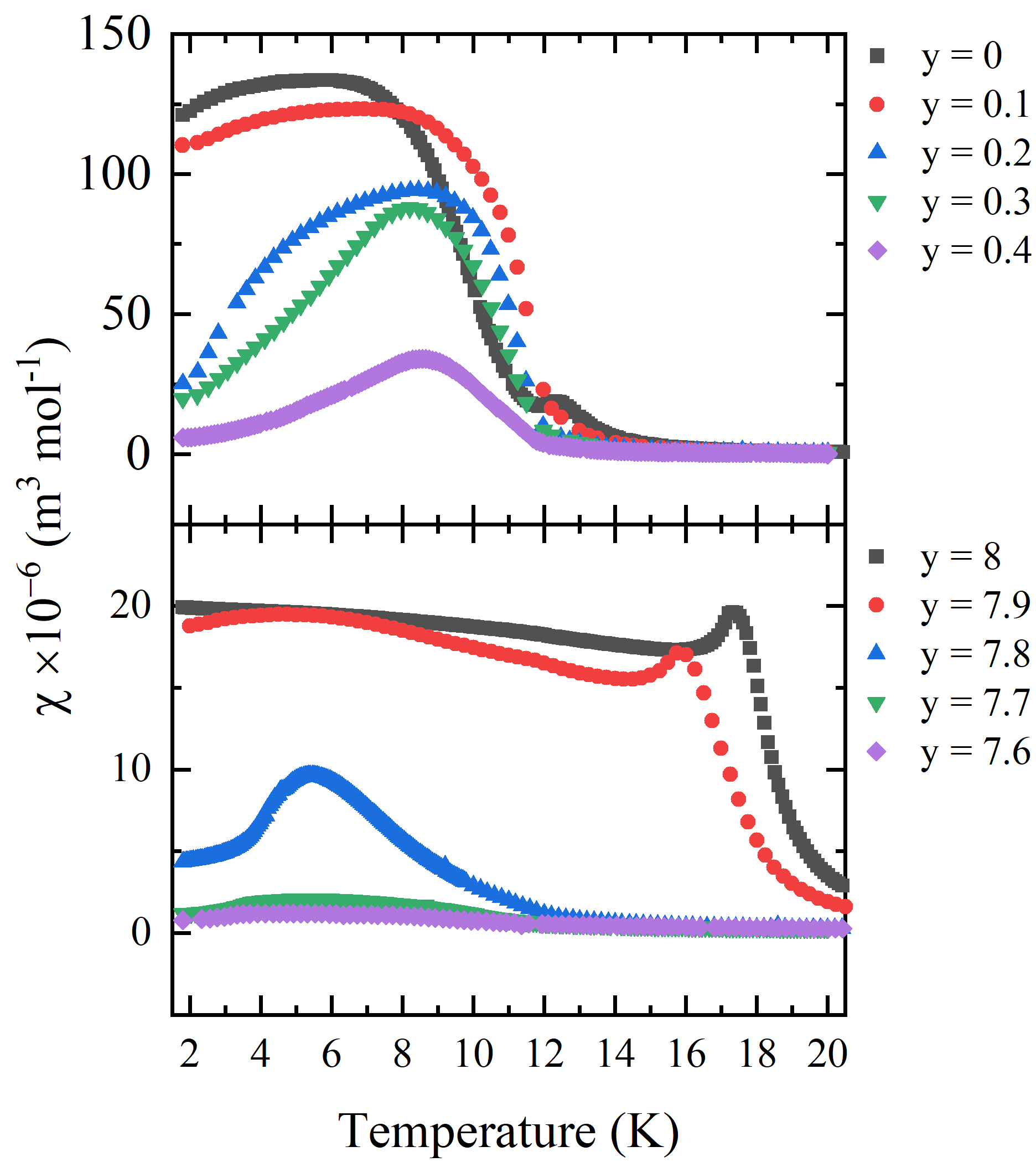}
\caption{Temperature dependence of the \textit{dc} magnetic susceptibility in an applied magnetic field of 10~mT collected after zero-field cooling for powders of the GaV$_4$S$_{8-y}$Se$_{y}$ family. Top: $0\leq y\leq 0.4$. Bottom: $7.6\leq y\leq 8$.}
\label{Fig:magnetization}
\end{figure}

The nature of magnetic ordering was examined using bulk magnetization measurements on the synthesized powders in the GaV$_4$S$_{8-y}$Se$_{y}$ series.
Figure~\ref{Fig:magnetization} shows the \textit{dc} magnetic susceptibility of the powders investigated.
The magnitude of the magnetization observed is significantly larger in \ce{GaV4S8} compared to \ce{GaV4Se8} due to the different magnetic ground states.
These parent compounds \ce{GaV4S8} and \ce{GaV4Se8}, undergo long-range ferromagnetic-like ordering at 13 and 18~K, respectively in agreement with literature~\cite{kezsmarki2015neel, bordacs2017equilibrium}.
A dramatic decrease is seen in the zero-field-cooled warming \textit{dc} susceptibility for both \ce{GaV4S8} and \ce{GaV4Se8} with increasing levels of substituents.
The addition of S to \ce{GaV4Se8} ($7.5\leq y\leq 8$) produces a more pronounced decrease in the magnitude of magnetization than the addition of Se to the \ce{GaV4S8} ($0\leq y\leq 0.5$).
As the magnetic phases that are stabilized change with substitution, it is difficult to pinpoint exactly where the paramagnetic to magnetic transition occurs.
The difference in the \textit{dc} susceptibility, shown in Fig.~\ref{Fig:magnetization}, between the $y=0$ and $0.1$ compositions is apparent as the smaller peak centered at 12.5~K in the $y=0$ compound has merged with the lower temperature transition.
Compositions with $0.2\leq y\leq 0.4$ then follow a trend of decreasing transition temperature and reduction of the bulk magnetization at lowest temperature.
Neutron diffraction measurements for $y=0.4$ indicate that the material contains only $\sim 30\%$ of the rhombohedral phase at 1.5~K.
In the Se rich materials, a dramatic change in the magnetization is observed between $y=7.9$ and $7.8$ indicating the changing nature of the magnetic state.
For $7.6\leq y\leq 7.8$ the magnetization decreases in magnitude with a broad transition beginning at 12~K with $y=7.8$ peaking at $\sim 5.7$~K due to the freezing of the spins at lower temperatures.
One of the likely reasons for the decrease in magnetic susceptibility is due to the substituent disrupting the magnetic exchange pathways.
The observation of the strong preferred substitution sites in these materials suggests that substitution of S in the V$_4X_4$ clusters has a greater effect on the exchange pathways than that of Se in the Ga$X_4$ tetrahedra.
Density functional theory results on substituted GaV$_4$S$_{8-y}$Se$_{y}$ from Ref.~\cite{hicken2020magnetism} support the evidence for the alteration of exchange pathways based on substitution at different sites.
These results are complementary to our previous studies \cite{stefancic2020establishing} and highlight the progression of the ordered magnetism of the parent compounds towards the spin glass-like state for compounds with $1\leq y\leq 7$.

\begin{figure}
\centering
\includegraphics[width=0.95\linewidth]{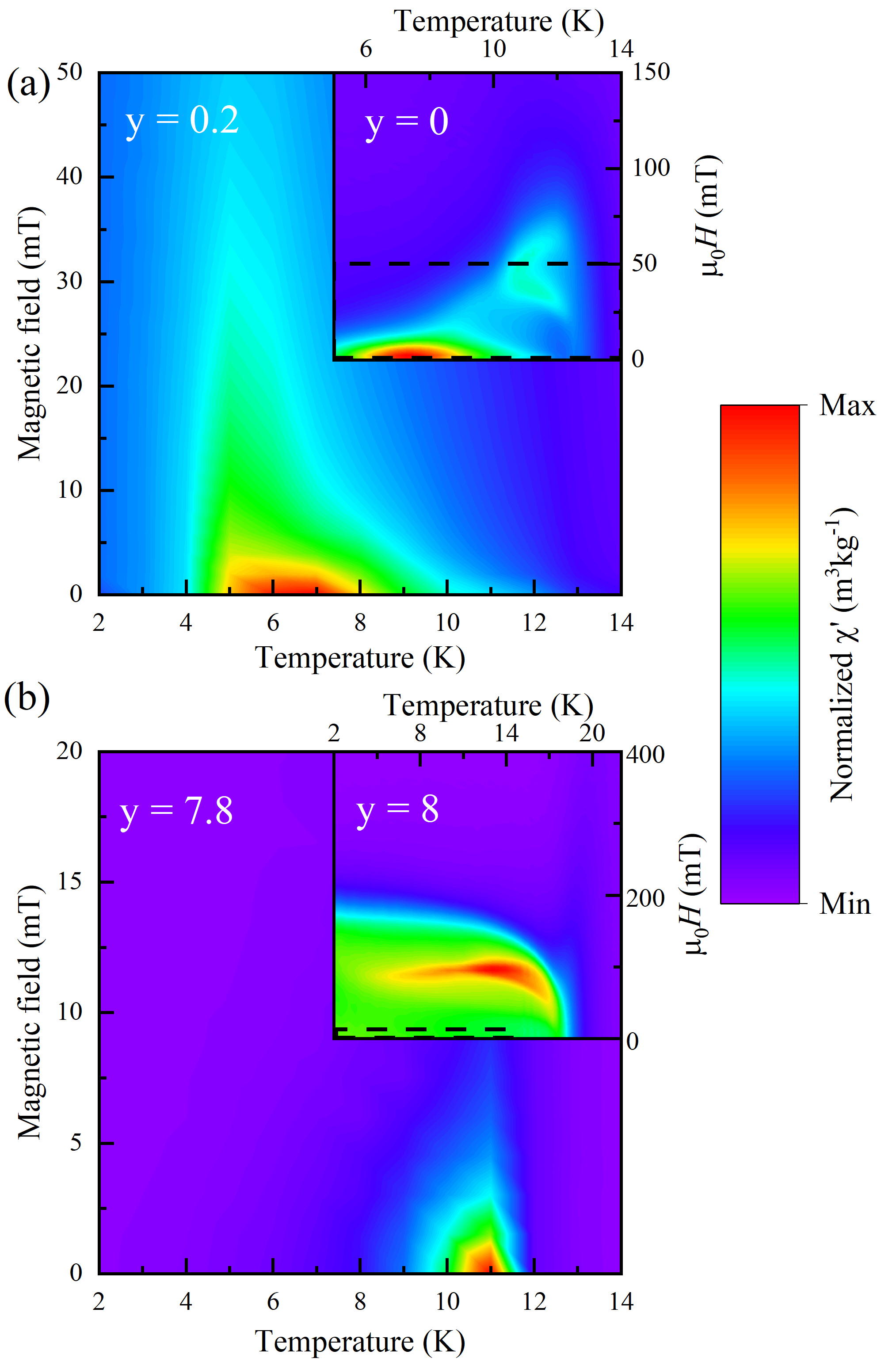}
\caption{Phase diagrams of the normalized real component $\chi'$, of the \textit{ac} susceptibility in an \textit{ac} field of 0.3~mT with a frequency of $113$~Hz as a function of temperature and \textit{dc} applied magnetic field for GaV$_4$S$_{8-y}$Se$_{y}$.
$\chi_{\mathrm{min}}=7.3\times~10^{-8}$ is consistent for all plots but $\chi'_{\mathrm{max}}$ is unique to each composition due to the varying orders of magnitude present.
(a) $y=0.2$ ($\chi'_{\mathrm{max}}=1.4\times~10^{-5}$) with $y=0$ inset ($\chi'_{\mathrm{max}}=7.7\times~10^{-6}$) and (b) $y=7.8$ ($\chi'_{\mathrm{max}}=2.57\times~10^{-4}$) with $y=8$ inset ($\chi'_{\mathrm{max}}=4.47\times~10^{-4}$).
Dashed lines in the insets show the temperature - field region plotted in the main panels.}
\label{Fig:AC}
\end{figure}

\begin{figure}
\centering
\includegraphics[width=0.9\linewidth]{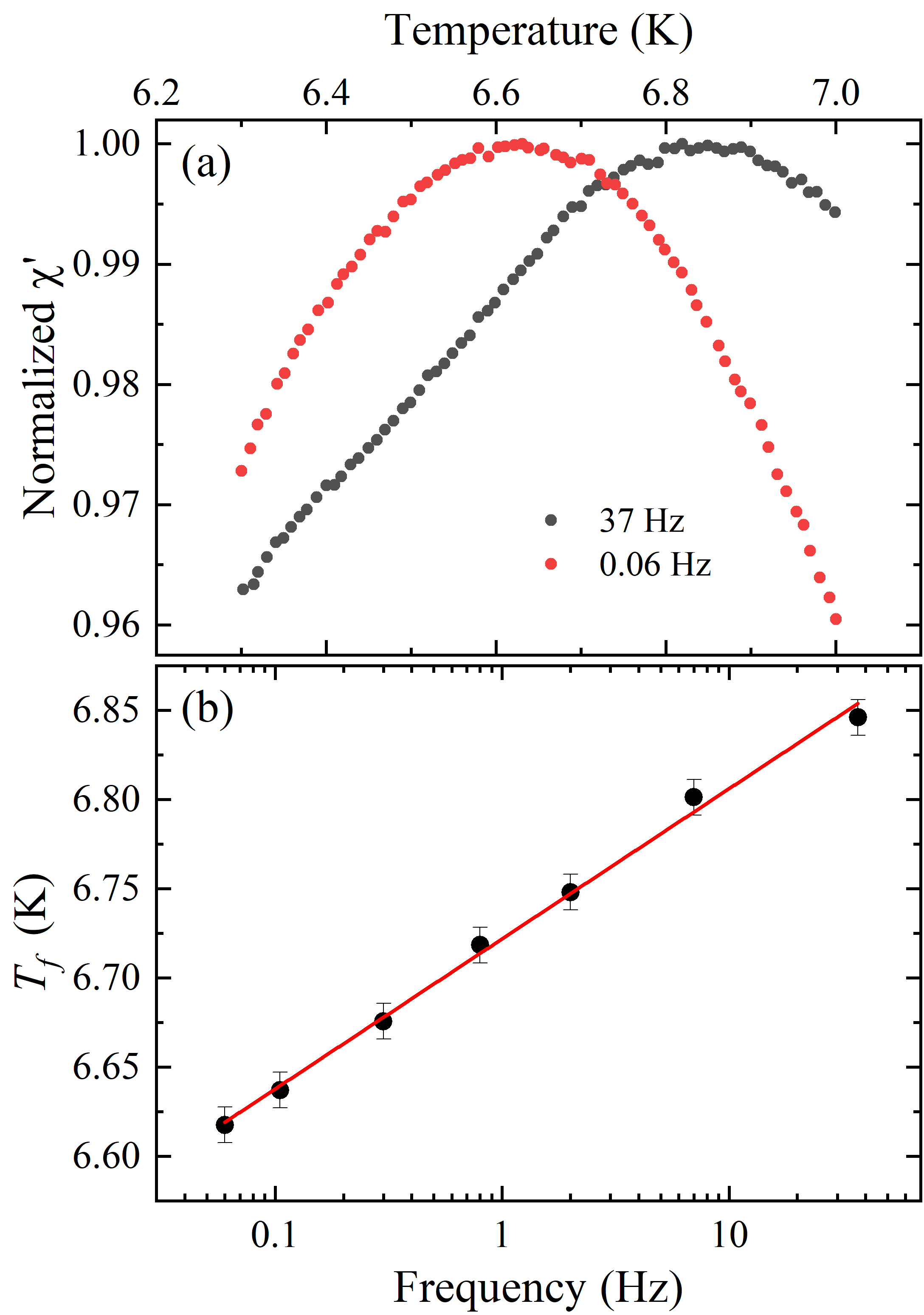}
\caption{(a) Temperature dependence of the normalized real component $\chi'$ of the \textit{ac} susceptibility for GaV$_4$S$_{0.2}$Se$_{7.8}$ $(y=7.8)$ spanning the freezing transition $T_f$.
The measurements were performed in zero applied \textit{dc} magnetic field and an \textit{ac} magnetic field of 0.3~mT amplitude for the frequencies of 37 and 0.06~Hz.
(b) Semi-logarithmic plot of $T_f$ ranging over multiple decades of frequency for GaV$_4$S$_{0.2}$Se$_{7.8}$.}
\label{Fig:AC_freq}
\end{figure}

The dynamics of the magnetism in these materials can be probed using \textit{ac} susceptibility.
Different magnetic phases have different characteristics in \textit{ac} susceptibility, therefore once a magnetic phase is identified, \textit{ac} susceptibility can be a powerful technique to examine the magnetic phase diagram \cite{bauer2012magnetic}.
It is important to note that due to the easy axis and easy plane anisotropy in these materials the angle of applied magnetic field with respect to the crystallographic axis can affect the magnetic phases stabilized and the fields at which they occur.
However, when measuring polycrystalline materials what is observed is effectively an average of all crystallographic orientations.
Examining the \textit{ac} susceptibility phase diagrams for $y=0.2$ and 7.8, shown in Fig.~\ref{Fig:AC} reveals a dramatic change in the topography of both phase diagrams compared to the pristine materials. 
In Fig.~\ref{Fig:AC}(a) the phase diagram of GaV$_4$S$_{7.8}$Se$_{0.2}$ can be compared to that observed for \ce{GaV4S8}, however, there has been a loss of some features. 
The clear pocket, centered at 12~K and 50~mT, associated with the existence of a skyrmion lattice for \ce{GaV4S8} can no longer be seen in $y=0.2$ material.
Figure~\ref{Fig:AC}(b) shows that for GaV$_4$S$_{0.2}$Se$_{7.8}$ there is a complete suppression of all of the features that are observed for \ce{GaV4Se8} with a peak seen at $\sim 11$~K in zero-field indicating the onset of magnetism ordering.
This onset can be seen at $\sim 11.5$~K in the \textit{dc} susceptibility shown in Fig.~\ref{Fig:magnetization}.

To further investigate the glassy-like transition observed for $y=7.8$ (shown in Fig.~\ref{Fig:magnetization}) we have carried out the frequency dependence of the \textit{ac} susceptibility around the freezing transition.
The measured \textit{ac} susceptibility for two different frequencies is shown in Fig.~\ref{Fig:AC_freq}(a) and a clear decrease in temperature of the peak can be seen as the frequency of the applied \textit{ac} magnetic field is decreased.
This indicates the presence of a glassy-like magnetic state in this material, and is consistent with our observations of a similar state in members of this family with higher levels of substitution \cite{stefancic2020establishing}.
In GaV$_4$S$_{0.2}$Se$_{7.8}$, the observed peak corresponds to a freezing temperature $T_f$ of $\sim 5.4$~K from the \textit{dc} susceptibility curve shown in Fig.~\ref{Fig:magnetization}.
Figure~\ref{Fig:AC_freq}(b) shows the frequency dependence of $T_f$. 
The shift of the freezing temperature per decade of frequency $\frac{\Delta T_f}{T_f \Delta(\log_{10}\nu)}$,
where $\nu$ is the frequency of the applied \textit{ac} magnetic field, is $\sim 0.016$ in good agreement with what is observed for \ce{GaV4S4S4} and canonical spin-glasses \cite{stefancic2020establishing, mydosh}.

The powder \textit{ac} susceptibility phase diagrams shown in Ref.~\cite{hicken2020magnetism} for $y=0$, 0.1, 7.9, and 8 contain features that can be associated with the skyrmion, cycloidal, and ferromagnetic-like magnetic phases in these materials.
The attribution of these features to magnetic phases in the substituted materials are supported by $\mu^+$SR measurements \cite{hicken2020magnetism}.
A comparison of the phase diagrams determined from \textit{ac} susceptibility data for the two parent compounds, would appear to suggest that for compositions of GaV$_4$S$_{8-y}$Se$_{y}$ with $0 \leq y<0.2$ and  $7.8 < y\leq 8$ skyrmions lattices could be stabilized.
However, for $0.2\leq y\leq 7.8$ the lack of these features suggests that skyrmions lattices with similar dynamics to the parent compounds may no longer exist.

The differences in the magnetic response of the substituted materials compared to the end compounds have two main possible sources.
The first, of local origin, is the preferred substitution modifying the exchange pathways and hence the global magnetism.
The preferential substitution of Se into the \ce{GaX4} tetrahedron affects the inter-cluster exchange whereas S into the \ce{V4X4} cluster affects the intra-cluster exchange.
The second source is the reduction of the structural distortion that can modify the nature of the exchange including the Dzyaloshinskii-Moriya interactions leading to the different magnetic phases present.
There is a smaller change to the distortion (as quantified by the $c/a$ ratio) but a larger change in the magnetism observed at the \ce{GaV4Se8} end of the family compared to the \ce{GaV4S8} end.
This suggests that an alteration of the exchange pathways brought about as a result of the preferred occupancy of S and Se in the \ce{V4X4} clusters and \ce{GaX4} tetrahedra, has a larger effect on the resulting magnetic properties than any changes caused by the structural distortion.

\section{\label{sec:Conclusions} Conclusions}
We have undertaken a detailed investigation of the structural and magnetic properties of the GaV$_4$S$_{8-y}$Se$_{y}$ family of materials for substitutions in the range $0\leq y\leq 0.5$ and $7.5\leq y\leq 8$, to gain a greater understanding of the effects of low levels of substitutions.
Powder neutron diffraction measurements show strong evidence for the first order nature of the structural phase transition in this family.
The addition of substituents to the parent compounds leads to a gradual shift of the structural phase transition to lower temperatures, combined with a reduction in the structural distortion until a critical amount of the substituents completely suppresses the transition. 
In the compositions which undergo the structural transition, the disorder effects lead to the observation of a striking region of coexistence of the cubic ($F\Bar{4}3m$) and rhombohedral ($R3m$) phases, in the vicinity of the structural phase transition.

Bulk magnetic property measurements show a decrease in the \textit{dc} magnetization with addition of substituents.
This process occurs faster at the Se end where the addition of S into the crystal lattice is seen to have a larger effect on the magnetic exchange pathways, due the preferential sites the S and Se occupy in the lattice.
Studies of the evolution of the \textit{ac} susceptibility to examine the magnetic phase diagrams for these materials show that the phase diagrams for substitution levels as low as $y=0.2$ and 7.8 are significantly altered from those observed in the parent skyrmion hosting materials.
This indicates that skyrmion lattices with similar dynamics to the parent compounds are present only in materials with less than $2.5\%$ substitution levels, i.e. $0 \leq y<0.2$ and  $7.8 < y\leq 8$.

We conclude that the substitution of S and Se even at very low levels $(2.5\%)$ brings about a large enough disruption to the crystal structure, and this combined with their preferential site occupancies is sufficient to alter the magnetic phases exhibited by these materials in comparison to the end compounds.
The fragility of the structure to any manipulation is strongly linked to the ability of these lacunar spinel structures to host skyrmion lattices.
Our findings on the effect of subtle structural distortions adds to the evidence that already exists from studies on similar lacunar spinels such as \ce{GaMoSe8} \cite{schueller2020structural}.
Studies such as this are vital to obtain a deeper understanding of the structure-magnetism correlations in the lacunar spinel family as well as in other families of skyrmion materials. 


\section*{Acknowledgments}
This work is financially supported by the EPSRC (EP/N032128/1).
We are grateful for the beamtime allocated to us at the Institut Laue-Langevin (ILL) under the experiment code 5-31-2684.
Data is available from the ILL at DOI: 10.5291/ILL-DATA.5-31-2684.
We would like to thank T. J. Hicken and T. Lancaster (Durham University) for their critical reading of the manuscript.

\end{document}